\documentclass[preprint,12pt]{elsarticle}
\usepackage{mathrsfs,ifthen}
\usepackage{amssymb}
\usepackage[usenames]{color}

\newboolean{pdf}
\ifthenelse{\isundefined{\pdfoutput}}   %are we running under pdflatex?
    {\setboolean{pdf}{false}}
    {\setboolean{pdf}{true}}

\usepackage{graphicx,subfigure}
\graphicspath{{./}{./Figs/}}

\ifpdf \DeclareGraphicsExtensions{.pdf,.jpg,.jpeg,.png}
\else
\DeclareGraphicsExtensions{.eps,.eps.gz,.jpg,.jpeg,.png}
 \fi

\usepackage{algorithm, algorithmic}

\usepackage{tabularx,multirow,hhline}

\journal{Mathematics and Computers in Simulation}

\begin{document}

\begin{frontmatter}


% use the tnoteref command within \title for footnotes;
% use the tnotetext command for theassociated footnote;
% use the fnref command within \author or \address for footnotes;
% use the fntext command for theassociated footnote;
% use the corref command within \author for corresponding author footnotes;
% use the cortext command for theassociated footnote;
% \author{Name\corref{cor1}\fnref{label2}}
% \address{Address\fnref{label3}}

\title{Complex Agent Networks explaining the HIV epidemic \\ among homosexual men in Amsterdam}


\author[UvA,NUDT]{Shan Mei\corref{cor1}}
\ead{A.MeiShan@uva.nl, MeiShan@nudt.edu.cn}

\author[UvA]{P.M.A Sloot}

\author[UvA]{Rick Quax}

\author[NUDT]{Yifan Zhu}
\author[NUDT]{Weiping Wang}

\cortext[cor1]{Corresponding author}

\address[UvA]{Computational Science, University of Amsterdam, Amsterdam, Netherlands}
\address[NUDT]{National University of Defense Technology, Changsha, P. R. China}

\begin{abstract}
% Text of abstract
Simulating the evolution of the Human Immunodeficiency Virus (HIV)
epidemic requires a detailed description of the population network,
especially for small populations in which individuals can be
represented in detail and accuracy. In this paper, we introduce the
concept of a Complex Agent Network(CAN) to model the HIV epidemics
by combining agent-based modelling and complex networks, in which
agents represent individuals that have sexual interactions. The
applicability of CANs is demonstrated by constructing and executing
a detailed HIV epidemic model for men who have sex with men (MSM) in
Amsterdam, including a distinction between steady and casual
relationships. We focus on MSM contacts because they play an
important role in HIV epidemics and have been tracked in Amsterdam
for a long time. Our experiments show good correspondence between
the historical data of the Amsterdam cohort and the simulation
results.
\end{abstract}

\begin{keyword}Multi-Agent Systems(MAS) \sep Complex Network(CN) \sep HIV
epidemics \sep scale-free \sep Men who have Sex with Men(MSM)

% \PACS 89.75.Fb \sep  89.75.Hc \sep  05.65.+b \sep 87.90.+y

% MSC codes here, in the form: \MSC code \sep code
% or \MSC[2008] code \sep code (2000 is the default)

\end{keyword}

\end{frontmatter}

\section{Introduction}
\label{Introduction}

Understanding the underlying dynamics in Human Immunodeficiency
Virus (HIV) epidemics is a crucial public health issue,
unfortunately however addressing specific problems in small
populations is difficult because individuals need to be modelled
with detailed social behavior. Traditional mathematical methods
greatly simplify both the disease dynamics and the population
networks, however extending them to more detailed models is
intractable.

% Furthermore, whether such methods apply to networks of small sizes
%is still an open question [ref].

%Moreover, the use of these methods becomes impractical for those in
%which highly detailed individuals are involved in small human
%groups.

% Traditional
%mathematical methods of simulating infectious diseases in
%populations greatly simplify both the disease dynamics and the
%population network, making their results questionable for more
%complex dynamics.

In particular, whether such methods also apply to networks of small
size, and thus to many real-world biological or epidemiological
applications, is still an open question\cite{Pautasso2008}. Many
networks of relevance to epidemiology may be of relatively small
size, among which social contact networks on sexually transmitted
diseases for a small group (e.g. men who have sex with men, namely
MSM, within a city or town) are representative.

Modelling the HIV epidemic is difficult because the true
incidence\footnote{\textbf{Incidence} is a measure of the risk of
developing some new condition within a specified period of time. In
this paper we use the \textbf{incidence rate} to denote the number
of new cases per unit of person-time at risk.} of the
HIV/AIDS-epidemic is uncertain since many people may be unaware of
their infection. Secondly, HIV progression has a very long
asymptomatic period which makes studies of the actual infection
spreading a very complicated task \cite{Sloot2008}. Finally, the
various routes of infection and the inhomogeneity of the involved
population pose additional challenges to understanding the
underlying knowledge of HIV epidemics.

Multi-agent systems (MAS) and complex networks (CN) are often used
separately to model and simulate epidemics; however, whether
existing models, which typically focus on large populations, can
address epidemics among small groups ($\sim 10^3\texttt{-}10^4$) are
still questionable and seemingly need further validation. In terms
of agent-based modelling, Teweldemedhin developed an agent-based
bottom-up modelling approach for estimating and predicting theX
spread of the HIV in a given population \cite{Teweldemedhin2005};
Xuan developed an extended Cellular Automata simulation model to
study the dynamical behavior of HIV/AIDS transmission by
incorporating heterogeneity into agents' behavior \cite{Xuan2008}.
In terms of complex network modelling, Bai discussed a sexual
network spreading model for HIV epidemics \cite{Bai2007}; Sloot
proposed a new way to model HIV infection spreading through the use
of dynamic complex networks, with the time evolution of the network
vertices modelled by a Markov process \cite{Sloot2008}.

In this paper we present the CAN approach for simulating epidemics
in small networks to great detail. The CAN is a hybrid approach in
which multi-agent systems and complex networks are the basic methods
of modelling epidemics on individual and population scales
respectively. Using the CAN approach, we simulate a relatively
detailed model of the HIV endemic among MSM in Amsterdam. This model
includes a distinction between steady and casual relationships,
which is regarded as an important aspect by others
\citep[e.g.][]{Xiridou2003}. We compare the results to the Amsterdam
Cohort Study (ACS) historical data.

This paper is organized as follows. In Section \ref{framework} we
review MASs and complex networks and introduce the CAN approach. We
describe the HIV epidemic model in detail in Section \ref{model} and
simulation implementation in Section \ref{Implementation}. In
Section \ref{application} we discuss a case study and its simulation
results.

%\todo{Do you see how the `implemention details' logically follows the intro of CAN? I suggest you switch sections III and IV.}
%
%\todo{I deleted the following from the proze TOC because it is specific; put it in the resp. section, not here:
%``...details of the dynamics of HIV epidemics involving
%progression of infection, construction of contact networks,
%propagation of HIV and demographic influences.''}

\section{Complex Agent Network}
\label{framework}
%The Complex Agent Network is proposed to support
%HIV epidemic study. In terms of agents, it deals with MAS M\&S
%issues covering microscopic individual and macroscopic collective
%behavior. We are aiming at combining both advantages of using
%Multi-Agent Systems and Complex Networks.

The CAN approach takes the MAS and CN as basic methods for modelling
and simulating epidemics on individual and population scales. Agents
contain specified personal information on an individual scale,
whilst complex networks emphasize the relationship dynamics among
the agents on a population scale. We review both concepts and the
combination of them in this section.

\subsection{Multi-agent Systems}

A MAS is a system composed of multiple interacting intelligent
agents, and can manifest self-organization and complex behavior even
when the individual strategies of all the agents are simple.
Heylighen defined self-organization as ``the spontaneous emergence
of global coherence out of the local interactions between initially
independent components'' \cite{Heylighen2001}.

%Multi-agent systems can be used to solve problems which are
%difficult or impossible for an individual agent or monolithic system
%to solve.

The agents in a multi-agent system have several important
characteristics \cite{Michael2002}: \emph{Autonomy}: the agents are
at least partially autonomous. \emph{Local views}: no agent has a
full global view of the system, or the system is too complex for an
agent to make practical use of such knowledge.
\emph{Decentralization}: there is no controlling agent (or the
system is effectively reduced to a reductionistic system)
\cite{Panait2005}.

Applied to HIV epidemics, a MAS represent a specified human
community and viruses propagate along social contacts. Each
individual has his own progression of infection and simple rules to
choose partners. Each agent only knows the information about himself
and his partners instead of epidemiological statistics. There is no
dominating agent in the community to control the spread of viruses.

%Therefore, understanding network dynamics of disease spreading gives
%insights in discovering and verifying basic principles and key
%pathways in epidemiology.

\subsection{Complex Networks}

%The complex network founds on the concept of network (graph) theory,
%but including probable larger amount of nodes and substantial
%non-trivial topological features.

The study of complex networks is inspired largely by the empirical
study of real-world networks such as computer networks and social
networks. A network is a set of items, which we call vertices, with
connections between them, called edges \cite{Newman2003}. In the
context of network theory, a complex network is a network with
non-trivial topological features that mostly do not occur in simple
networks such as lattices.

The complex network theory aims to do three things\cite{Newman2003}.
Firstly, it aims to find and highlight statistical properties, such
as path lengths and degree distributions, that characterize the
structure and behavior of networked systems, and to suggest
appropriate ways to measure these properties. Secondly, it aims to
create models of networks that can help us to understand the meaning
of these properties, how they came to be as they are, and how they
interact with one another. Finally, it aims to predict what the
behavior of networked systems will be, on the basis of measured
structural properties and the local rules governing individual
vertices.

For HIV epidemics, an at-risk population can be depicted as a
network with vertices representing persons and edges representing
social contacts. Over these relationships, viruses are propagated
from one individual to another and thus give rise to the epidemic
prevalence. Findings in sociology or epidemiology suggest that
social contact networks are scale-free networks
\cite{Schneeberger2004}. Therefore we create networks with vertices
following power-law degree distributions, and then simulate HIV
propagation by evolving networks temporally and spatially.

%The statistical properties of network, such as degree distribution,
%assortativity or mixing patterns, and clustering coefficient etc.,
%provide insights into evaluating both network static structure and
%evolving dynamics. So that they can also assist evaluating statistical
%significance of simulation results and then analyzing influences of
%different factors on the epidemic prevalence.

% Our work enables more detailed investigation into epidemics.

\subsection{Complex Agent Networks}% for HIV epidemics}
% function: how + why to combine CNs and MASs for better simulations?

The benefits of combining a MAS and a CN is twofold.

Firstly, the edge forming and rewiring mechanism of a complex
network with a non-trivial topology can regulate the interactions
among agents in a MAS. In this way we can keep these interactions
from being either purely regular or purely random as we usually see
in a MAS. Furthermore, the recent development of statistical methods
for quantifying complex networks is a way to analyze the influence
of network topologies on epidemics \citep[e.g.][]{Goodreau2006}.

\begin{figure}
%\begin{center}
\includegraphics[width=200 bp]{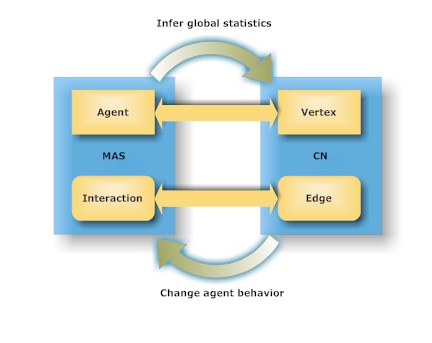}
\caption{Complex Agent Networks} \label{Framework}
%\end{center}
\end{figure}

Secondly, local agent information can be combined to infer global
statistics which can be fed back into the simulation dynamically,
resembling real-world epidemics more closely. For example, if the
HIV prevalence becomes very high, individuals may be less prone to
unsafe sexual interactions with other individuals whose serostatus
they do not know.\footnote{Here we assume that individuals somehow
find out such global statistics, e.g. through media.} This
interaction of both the individual and population levels is depicted
in Figure \ref{Framework}.

%; we have not yet implemented the ``Changes behavior'' feedback in
%our simulator.

% Secondly, a MAS can provide arbitrarily detailed description of
% autonomous individuals and a flexible scheduling mechanism.

%Thus
%epidemic-related parameters can be distilled from agents and the
%values of these parameters can be fed into networks dynamically. In
%addition, the scheduling mechanism of a MAS can drive the dynamic
%evolution of networks by scheduling agents then providing updated
%information of agents regularly.

% The status information of agents can be combined to calculate epidemic-related
% statistics, which can be fed back into the simulation dynamically. For example,
% if the HIV prevalence becomes relatively high, agents may be less prone to
% risky sexual interaction. We have not yet implemented this.

To combine the MAS and CN we map agents and their interactions in a
MAS to vertices and edges in a CN, respectively, as shown in Fig.
\ref{Framework}. Agents provide personal heterogeneity and local
infection progression on an individual level; networks benefit from
continuously distilling values of epidemic-related parameters from
individuals and then mimic the virus propagation on a population
level.

%Here,
%individuals are modelled by agents in a MAS and the interactions
%between individuals are regulated by a CN.

\subsection{Application of CAN to HIV epidemics}
%function: how does CAN fit into the big picture of simulating epidemics?

\begin{figure}
%\begin{center}
\includegraphics[width=400 bp]{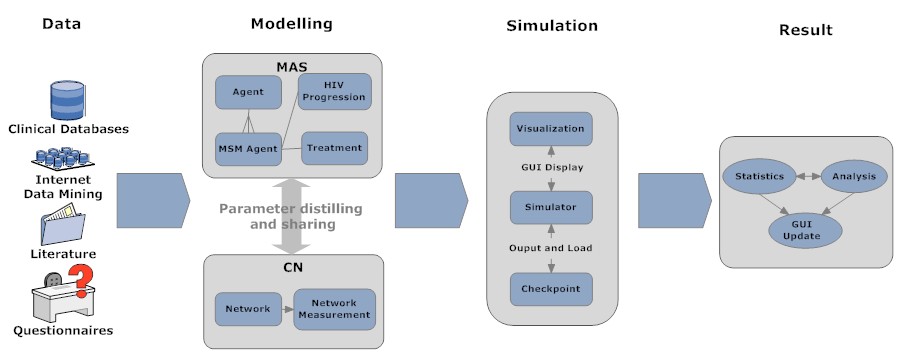}
\caption{Application reference procedure of Complex Agent Network}
\label{ArchitectureFig}
%\end{center}
\end{figure}

Fig. \ref{ArchitectureFig} shows an application reference procedure
of applying the CAN approach to HIV epidemics, which consists of
four steps, i.e. data collection, modelling, simulation and result
manipulation.

The CAN-based HIV epidemic studies commence with data collection
from e.g. \emph{Clinical Databases}, \emph{Internet Data Mining},
\emph{Literature} and \emph{Questionnaires}.

Fed with the data, models are constructed by utilizing the features
of both the MAS and CN, which are depicted by two composite blocks.
The ``{\emph{MAS}}'' composite block deals with societal individuals
and their infection details. An \emph{Agent} stands for an
individual and supports specialized inheritance. As shown in Fig.
\ref{ArchitectureFig}, an inherited block \emph{MSM Agent} manages
MSM' infection properties and status transitions by using \emph{HIV
Progression} and \emph{Treatment} blocks. A \emph{MSM Agent} also
support inferring values of parameters for a ``{\emph{CN}}''. The
\emph{Treatment} block considers factors related to HIV ART or
HAART\footnote{\textbf{ARV}, Anti-Retroviral Therapy;
\textbf{HAART}, Highly Active Anti-Retroviral Therapy, is the
combination of at least two different classes of antiretroviral
drugs.} (e.g. drug types, patients' adherence and drug resistances
\cite{SlootGrid2008}). The ``{\emph{CN}}'' composite block
manipulates network initializing, reshuffling and measuring through
the \emph{Network} and \emph{Network Measurement} blocks. In
addition, the ``{\emph{CN}}'' provides network rendering support for
the \emph{Visualization} block of simulation.

The \emph{Visualization} and \emph{Checkpoint} blocks can be invoked
by the \emph{Simulator} according to specified purposes, to enhance
interactive exploration of simulations.

The \emph{Statistics} and \emph{Analysis} blocks are applied to
simulation results and then enforce the \emph{GUI Update} of network
rendering and parameter plotting.

\section{Modelling HIV Epidemics}
\label{model}
%In this section, a model is presented to illustrate
%the application of CAN framework.

\subsection{Progression of HIV infection }
\label{infection progression} The progression of HIV infection from
seropositive to AIDS can broadly be divided into three stages:
primary infection (PI), asymptomatic period (AP) and AIDS. Although
there are much variations amongst patients, on average,
transmissibility is the highest during the primary infection which
is believed to be associated with high plasma HIV RNA levels and
continued risky sexual activities \cite{Koopman1997,Baggaley2005}.

Correspondingly each agent representing an at-risk individual,
possesses a disease manager to conduct its HIV infection progression
temporally. The manager conducts an irreversible ``negative
$\rightharpoonup$ PI $\rightharpoonup$ AP $\rightharpoonup$ AIDS''
order. Note that, we exclude the case that an individual with AIDS
can with the right treatment go back to the asymptomatic phase,
because in general patients are most likely diagnosed and treated
before reaching the AIDS stage.

%As shown in Figure \ref{HIVProgression}, the four stages are denoted
%as classes inherited from an abstract \emph{HIVStage} interface.
%
%
%\begin{figure}
%%\begin{center}
%\includegraphics[width=400 bp]{HIVProgression-i}
%\caption{HIV.Epidemics.PersonAgent.HIVProgression package}
%\label{HIVProgression}
%%\end{center}
%\end{figure}

\subsection{Construction of contact networks}
\label{Construction}

HIV is transmitted mainly through sexual contacts, intravenous drug
uses (IDU), mother-to-child transmissions and transfusion of
contaminated blood products. In this paper we focus on MSM contacts
which play an important role in HIV epidemics and have been tracked
in Amsterdam for a long time
\citep[e.g.][]{ACS2006,ACSOverview2006}. We model these contacts as
an undirected, scale-free complex network.

The sexual contact network starts with $N_{0}$ MSM, each of which is
assumed to be sexually active with an age between 15 and 65 based on
the age limits of the ACS samples \cite{ACSOverview2006}. The
network is assumed to be scale-free
\cite{Schneeberger2004,Reed2006}, whilst the degree $k$ of each
vertex follows a degree distribution $p_{k}\sim k^{-\gamma}$ to
represent his yearly number of partners. Let us choose
\begin{equation}\label{Pk}
p_{k}= \left\{
    \begin{array}{ll}
        Ck^{-\gamma} & \textrm{for $k \geq 1$}\\
          p_{0} & \textrm{for $k=0$}
\end{array} \right.
\end{equation}
where $\gamma$ and $p_{0}$ are constants and $C$ is a normalizing
factor.

After obtaining the degree sequence of vertices, edges(partnerships)
are formed by using the so-called configuration model
\cite{Newman2003}. A degree sequence is a set of $n$ values of the
degrees $k_{i}$ of vertices $i = 1,..., n$, from the given power-law
distribution. According to the configuration model, the probability
that a vertex is chosen as an end of a random edge is proportional
to $kp_{k}$.

Different from most previous studies on epidemics in static
networks, our networks evolve with simulation time steps, which is
driven by the HIV epidemic dynamics depicted in Section
\ref{propagation} and \ref{demographic}. Further implementation will
be discussed in Section \ref{operators}.

\subsection{HIV propagation in networks}
\label{propagation}

The HIV propagation in networks can become considerably difficult
when more details are taken into account. We provide a formula of
calculating the transmission probability, considering risk behavior,
treatments, (steady or casual) partnerships and transmissibility at
different infection stages.

Similar to the commonly used Susceptible-Infected-Removed (SIR)
epidemiological model, our model sets individuals in three discrete
states, i.e. susceptible, infected or removed (caused by deaths).
The infected state comprises primary infection, asymptomatic period
and AIDS stages (see Section \ref{infection progression}) which are
related with different statistical transmission probability ($TP$)
per action. These $TP$ per action are represented by
$TP_{\scriptsize{\textrm{PI}}}$, $TP_{\scriptsize{\textrm{AP}}}$ and
$TP_{\scriptsize{\textrm{AIDS}}}$ for PI, AP and AIDS stages,
respectively. Individuals at stage AIDS are assumed to have no
contacts, so that $TP_{\scriptsize{\textrm{AIDS}}}=0$.

Each individual $V_{i}$ has $k_{i}$ partners, among whom at most one
steady partner is allowed. This steady partnership lasts for years
based on statistical data. With the presence of a steady partner, an
individual is likely to reduce his risk behavior with all other
casual partners, and this possible reduction is represented as a
partnership factor. Two steady partners have many actions per year,
while two casual partners have one action per year.

For a vertex $V_{i}$, the statistical transmission probability per
action is denoted by $TP_{i}$. Then this transmission probability
per action is adjusted according to the stage an individual is
currently at, and the treatment reduction factor $F_{T_{i}}$ which
is derived from treatment effectiveness.

For an edge $E_{ij}$ as a partnership connecting $V_{i}$ and
$V_{j}$, the risk behavior factor is denoted by $F_{R_{ij}}$, which
is related to the partners' attitude towards risk; the number of
actions per year over this partnership is denoted by $N_{A_{ij}}$
($N_{A_{ij}}=N_{A_{ji}}, \forall i,j$).

In addition, the partnership factor denoted by $F_{P_{j}}$ is
subjected to whether the individual $V_{j}$ has a steady partner,
and it will influence all incident casual partnerships.

Then the transmission probability from $V_{i}$ to $V_{j}$ per action
is given based on parameters aforementioned
\begin{equation}\label{PTAij}
P_{TA_{ij}}=F_{P_{j}}*F_{R_{ij}}*(F_{T_{i}}*TP_{i})
\end{equation}

Because the PI stage lasts around 3 months (0.25 year) and falls
short of one time step of a year, we divide
$TP_{\scriptsize{\textrm{PI}}}$ into
$TP_{\scriptsize{\textrm{PI}},1}$ and
$TP_{\scriptsize{\textrm{PI}},2}$ as two transmissibility at PI
stage. Hence if $V_{i}$ is at the PI stage, we also divide
$P_{TA_{ij}}$ into $P_{TA_{ij},1}$ and $P_{TA_{ij},2}$, and
$N_{A_{ij}}$ into $N_{A_{ij},1}$ and $N_{A_{ij},2}$. Then

$P_{TA_{ij},1}=F_{P_{j}}*F_{R_{ij}}*(F_{T_{i}}*TP_{i,1})$,

$P_{TA_{ij},2}=F_{P_{j}}*F_{R_{ij}}*(F_{T_{i}}*TP_{i,2})$.

The transmission probability from $V_{i}$ to $V_{j}$ per year is

\begin{equation}\label{PTYij}
P_{TY_{ij}}= \left\{
    \begin{array}{ll}
        0 & \textrm{HIV negative  $V_{i}$} \\
        0.25*P_{TA_{ij},1}+0.75*P_{TA_{ij},2} & \textrm{PI $V_{i}$ and casual $E_{ij}$ }\\
        1-(1-P_{TA_{ij},1})^{N_{A_{ij},1}}*(1-P_{TA_{ij},2})^{N_{A_{ij},2}} & \textrm{PI $V_{i}$ and steady $E_{ij}$}\\
        P_{TA_{ij}} & \textrm{AP $V_{i}$ and casual $E_{ij}$} \\
        1-(1-P_{TA_{ij}})^{N_{A_{ij}}} & \textrm{AP $V_{i}$ and steady $E_{ij}$} \\
        0 & \textrm{AIDS $V_{i}$}
\end{array} \right.
\end{equation}

The probability of $V_{i}$ infected by positive partners per year is
\begin{equation}\label{PTYi}
P_{TY_{i}}=1-\prod^{k_{i}}_{m=1}(1-P_{TY_{mi}})
\end{equation}

where $k_{i}$ is the number of partners of $V_{i}$.

Note that $TP_{i}$, $F_{T_{i}}$ and $F_{P_{i}}$ are varying with the
status of $V_{i}$ and time, and so are $F_{R_{ij}}$ and $N_{A_{ij}}$
with the status of $E_{ij}$ and time. Therefore each susceptible
vertex gets infected with a probability given by Equation \ref{PTYi}
at each time step, which contributes to the dynamics of HIV
propagation.

\subsection{Demographic influences}
\label{demographic}

In this section, we discuss the demographic principles that regulate
the conservation or deletion (substitution) of individuals and
relationships.

The vertices in networks are initialized randomly with an age from a
range of 15 to 65, according to the age limits of the ACS samples
\cite{ACSOverview2006}. Each edge is initialized with an expected
duration according to the statistical real-world partnership
duration. At each time step, an individual gets an increment of age
and a partnership gets an increment of duration, mimicking the
reality.

If one of the following three conditions applies, then an individual
will be removed from a network:

(1) The individual's age exceeds 65, the upper age limit of the ACS
samples.

(2) The infection has developed to the AIDS stage (the individual
will be removed at the \emph{next} time step).

(3) In case of geographic migration (a fraction of randomly selected
individuals will be removed from a network. This fraction is left as
tunable, e.g. 1\% assumed in our model.).

Susceptible individuals will be added to the network as substitutes
for the removed individuals at each time step, to keep the
population size constant. All these newly added individuals will
join the epidemic contact network according to the principles given
in Section \ref{Construction}.

An edge will be kept for the \emph{next} time step if:

(1) It is a steady partnership and has not expired.

(2) A previous casual partnership is likely to be conserved with a
probability (e.g. 0.2 assumed in our model).

%\begin{algorithm}[H]
%\SetLine
% \KwIn{
% $Amateur^{+}$: Amateur-tagging images tagged with a specific concept,\\
% $Amateur^{-}$: Randomly selected Flickr images with $Flickr^{+}$ excluded,\\
% $GT$: Ground truth of the concept, created by expert
% tagging,\\
% $min$, $max$: minimum and maximum positive examples for training,\\
% $step$: example sampling interval,\\
% $max\_round$: sampling frequency.
% }
% \KwOut{Ranked $Flickr^{+}$ and $Flickr^{-}$.}
%
%\ForEach{$n$ in $Range[min, max, step]$}
% {
% \ForEach{round $i$ in $Range[1,max\_round]$ }
%  {
%   $trainset=RandSample(n, Flickr^{+})\cup RandSample(10\times n,
%   Flickr^{-})$;\\
%   $classifier=Learning(trainset)$;\\
%   $AP=Evaluation(classifier, GT)$;\\
%   \ForEach{example $x$ in trainset}
%    {
%       $AP_x=AP_x \cup \{AP\}$
%    }
%  }
% }
%
%\ForEach{example $x$ in $(Flickr^{+} \cup Flickr^{-})$}
% {
%   $Score(x)=\frac{\sum_{y \in AP_x}y}{|AP_x|}$
% }
%Rank $x \in Flickr^{+}$ according to $Score(x)$ in descend order\\
%Rank $x \in Flickr^{-}$ according to $Score(x)$ in descend order
% \caption{\small{``Good example'' Estimation by AP voting}}
%\label{alg:apvoting}
%\end{algorithm}

\section{Simulation Implementation}

\label{Implementation}

We implement specific functionalities of the HIV epidemics
simulation, such as an advanced scheduling mechanism and interactive
visualization, using the JUNG\cite{Madadhain-JUNG} and
MASON\cite{MASON2004} libraries.

\subsection{Functionality}\label{Functionality}
The specific functionalities of the CAN simulation contains the
following components:

(1) Model interpretation

The model described in Section \ref{model} is interpreted into an
executable simulation supporting the population and individual
scales.

On a population scale, the formation and reshuffle of the edges are
driven by the dynamics of the networks. An edge is formed based on
the degrees and local status of two potential partners, as well as
the (steady or casual) type and duration of their partnership.

On an individual scale, heterogenized agents are described with
different HIV progression stages (Susceptible, PI, AP and AIDS)
which are influenced per time step by the disease progression, risk
behavior and treatment.

Furthermore, the interpreted executable simulation consists of
modules which we call operators (see Section \ref{operators}). We
embed these operators into either agents or networks, and then
schedule them using a simulator.

(2) Simulator

We implement a flexible scheduling mechanism, interactive
visualization and a checkpoint mechanism into the simulator. The
scheduling mechanism is double-precision discrete-event-and-priority
driven and supports complicated user interactions with the
simulator. This gives support to interactive exploration of the
running simulations and, for instance, allows checkpoint/restart of
long running simulations.

%Based on the above features, simulations can be executed in
%combinations. For example, we can start/resume a simulation from
%loading a previous checkpoint file generated in
%switched-off-visualization and multi-job running mode (for
%computational performance) to obtain potential megascopic
%significances by observing its visualization display.

(3) Statistics and analysis

We can take specific epidemic-related statistics and analyze their
results, besides general statistical properties e.g. path lengths.
For example, we take incidence per person-year as a measure of HIV
epidemics in Section \ref{application}, by dividing the number of
newly infected persons every year by the population size.

\subsection{Operators}\label{operators}
The simulation operators with their annotated priorities are shown
in Table \ref{operators table}. These operators are invoked at each
time step to evolve a network $N[t]=\left<V[t], E[t]\right>$ where
$t$ is the time step.

\newcommand{\PBS}[1]{\let\temp=\\#1\let\\=\temp}
\newcolumntype{R}[1]{>{\PBS\raggedright\hspace{0pt}}m{#1}}
\newcolumntype{L}[1]{>{\PBS\raggedleft\hspace{0pt}}m{#1}}
\renewcommand{\tabularxcolumn}[1]{>{\PBS\raggedleft\hspace{0pt}}m{#1}}

\begin{table}
\centering \caption{Simulation operators} \label{operators table}
\newlength{\LL}
\settowidth{\LL}{Scale}
\begin{tabular}{R{2.6cm}R{2cm}R{1.7cm}R{6cm}}\hline
%\begin{tabularx}{\linewidth}{|l|*{4}{>{\small}X|}}

\textbf{Operator} & \textbf{Scale} & \textbf{Priority} &
{\textbf{Description}}\\\hline

{Infection Operator} & {Population} & {0} & {Infect susceptible
individuals and propagate HIV.} \\\hline

{Local Progression Operator} & {Individual} & {1} & {Advance HIV
disease progression of individuals.}
\\\hline

{Demographic-Reshuffling Operator} & {Population} & {2} & {Enforce
demographic and reshuffling influences on network evolutions.}
\\\hline

{Statistics and Visualization} & {Population} & {3} & {Statistics
approaches and visualization updates can be added and configured
according to pre-specified outputting rules.}\\\hline

%\end{tabularx}
\end{tabular}
\end{table}

The \textit{Infection Operator} basically deals with virus
propagation on a population scale. For simplification, we
disseminate its calculation to individuals by using Equations
\ref{PTAij}-\ref{PTYi} in Section \ref{propagation}. Algorithm
\ref{Infection operator} gives the details.

The \textit{Local Progression Operator} and \textit{Demographic -
Reshuffling Operator} are presented in Algorithms \ref{Progression
operator} and \ref{Demographic operator} respectively.

%Note that, for clarity, all operators are depicted with their own
%loops over vertices and edges, but in fact these loops are nested
%and combined in implementation to obtain a higher computational
%performance.

\begin{algorithm}
\small \caption{Infection Operator} \label{Infection operator}
\begin{algorithmic}
\STATE \textbf{Input}: Network $N=\left<V, E\right>$
 \STATE \textbf{Output}: The evolved network $N$ with some vertices newly infected.
% \STATE \textbf{Parameter}: neighbor number $k$
\STATE
 \STATE Infect susceptible vertices.
  \FOR {$V_{i} \in V$}
    \IF {$V_{i}$ is susceptible}
       \STATE Judge whether $V_{i}$ has any steady partnership
       \STATE Calculate $P_{TY_{i}}$ using Equations \ref{PTAij}-\ref{PTYi}
       \IF {a generated random number $<P_{TY_{i}}$}
           \STATE Set the infection stage of $V_{i}$ to Primary Infection.
       \ENDIF
    \ENDIF
  \ENDFOR
 \end{algorithmic}
\end{algorithm}

\begin{algorithm}
\small \caption{Local Progression Operator} \label{Progression
operator}
\begin{algorithmic}
\STATE \textbf{Input}: Network $N=\left<V, E\right>$
 \STATE \textbf{Output}: The evolved network $N$.
% \STATE \textbf{Parameter}: neighbor number $k$
\STATE
 \STATE  Make vertices and edges progress.
  \FOR {$V_{i} \in V$}
    \STATE $V_{i}$.Age++
     \IF {$V_{i}$ is infected}
        \IF {$V_{i}$.CurrentStage.LastingPeriod $<$
            \STATE $V_{i}$.CurrentStage.ExpectedDuration}
            \STATE $V_{i}$.CurrentStage.LastingPeriod++
        \ELSE
            \STATE $V_{i}$ progresses to next stage i.e. PI$\rightarrow$AP or AP$\rightarrow$AIDS.
        \ENDIF
    \ENDIF
  \ENDFOR
   \STATE
  \FOR {$E_{i} \in E$}
    \IF {$E_{i}$ is a steady partnership}
        \IF {$E_{i}$.LastingPeriod$ < E_{i}$.ExpectedDuration}
            \STATE $E_{i}$.LastingPeriod++
        \ENDIF
   \ENDIF
  \ENDFOR
 \end{algorithmic}
\end{algorithm}

\begin{algorithm}
\small \caption{Demographic-Reshuffling Operator} \label{Demographic
operator}
\begin{algorithmic}
\STATE \textbf{Input}: Network $N=\left<V, E\right>$
 \STATE \textbf{Output}: The evolved network $N$.
% \STATE \textbf{Parameter}: neighbor number $k$
\STATE
 \STATE  Evolve the network demographically.
  \FOR {$V_{i} \in V$}
    \STATE Remove $V_{i}$ according to the rules in Section \ref{demographic}.
  \ENDFOR
  \STATE Add new susceptible vertices to substitute removed ones.
   \STATE
  \FOR {$E_{i} \in E$}
   \STATE Remove $E_{i}$ according to the rules in Section \ref{demographic}.
  \ENDFOR
  \STATE Reshuffle all edges according to the rules in Section \ref{Construction}.
 \end{algorithmic}
\end{algorithm}

\subsection{Simulation scheduling}

We design and implement a flexible scheduling mechanism with three
main advantages.

Firstly, a double-precision discrete-event based scheduler is
available. With this we can schedule special events, such as the Gay
Parade Day\footnote{The \textbf{Gay Parade Day} is the national
celebration day on the first weekend in August in the Netherlands.}
parties for MSM happening at fractional time steps.

Secondly, all agents and operators can be registered with predefined
priorities. Vertices (agents) and edges (partnerships) are scheduled
firstly, (demographical and reshuffling) dynamics of networks
secondly, and statistics and GUI updates finally. In this way, the
infection and local progression drives the dynamic evolution of
networks and the collection of statistics.

Finally, the simulator can schedule, for special purposes, vertices
and edges at predefined intervals. For example, we can collect
incidence statistics every 2 years instead of 1 year.

\subsection{Visualization}
The visualization of contact networks can present an overview of the
studied population, and explain details about a selected vertex or
edge. The visualization serves the following purposes:

(1) To represent individuals in a population and (steady or casual)
relationships.

%(2) The visualization figure can be configured to update per one or
%several simulation time steps according to diverse application goals
%(e.g. more frequent updates versus higher computation performance).

(2) To arrange and display vertices and edges, in order to study the
sorting, grouping and correlation of vertices and edges.

(3) To present detailed properties of selected vertices or edges.

\begin{figure}
%\centering
 \caption{\textbf{Initialization(a)} and
\textbf{After reshuffling(b)} in the Amsterdam MSM simulation (The
\textit{green}, \textit{red}, \textit{blue} and \textit{black}
circles represent the susceptible group, the primary infection
group, the asymptomatic group and the AIDS group, respectively;
\textit{Solid} and \textit{dotted} lines represent steady
partnerships and casual partnerships, respectively; \textit{small},
\textit{medium} and \textit{large} circles represent vertices with a
degree less than 50, between 50 and 100, and greater than 100,
respectively.)}

\label{AmsterdamMSM} \subfigure[Initialization]{
\includegraphics[width=200 bp]{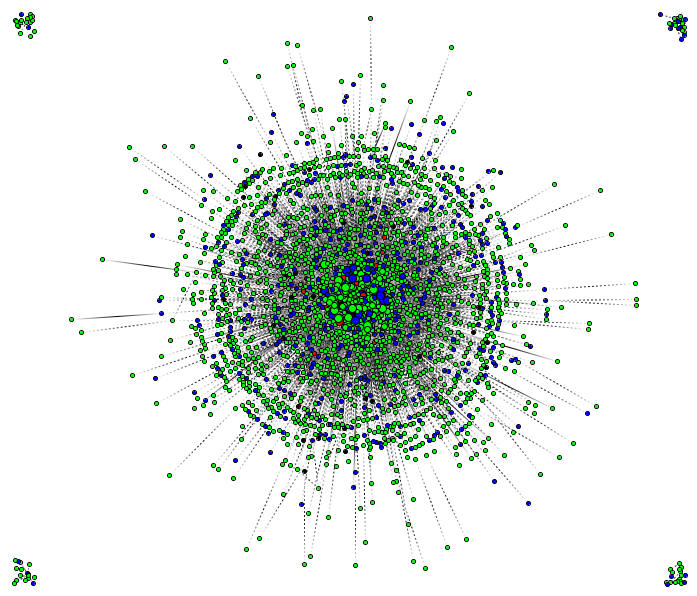}}
\subfigure[After reshuffling]{
\includegraphics[width=200 bp]{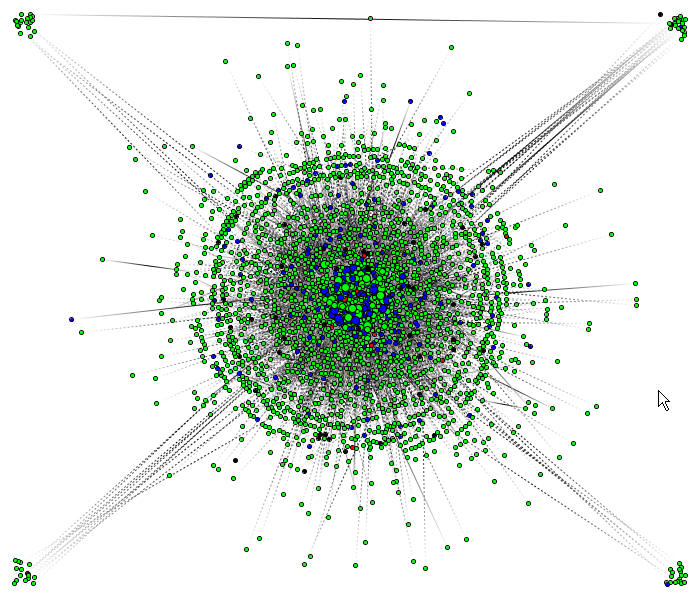}}
\end{figure}

Fig. \ref{AmsterdamMSM} shows the visualization of the network after
initialization and reshuffling.

\section{Case Study: the HIV epidemic in the Amsterdam MSM cohort}
\label{application} We take the HIV epidemic among the Amsterdam MSM
cohort as a case study.

\subsection{Amsterdam cohort study}

The Amsterdam Cohort Study (ACS) of HIV infection and AIDS among MSM
was initiated in 1984. According to the report \cite{ACS2006}, 2299
MSM have been involved in the ACS until 2006. Of the 2299 MSM, 571
were HIV-positive at study entry and 192 seroconverted during
follow-up. Yearly HIV inicidence for MSM over calendar years is
reported in \cite{ACSOverview2006} to indicate the HIV prevalence
trend among MSM.

\subsection{Parameters and their values}

\subsubsection{On a population scale}
Societal-related parameters for individuals and network-related
initialization are discussed.

Schneeberger concluded in \cite{Schneeberger2004} that MSM
population follows a power-law degree distribution with a value of
$\gamma$ in the interval between 1.5 and 2. We assume $\gamma = 1.6$
and a maximum degree $k_{max}=200$. For initialization, the network
size is $N_{0}=2299$, and the fraction of vertices with a degree of
0 is set to be $p_{0}=0.01$ because we assume a small portion of
people in a population not having any contact.

% in the range of 1.5 to 2.0

Steady relationships among MSM have a higher contribution to HIV
incidence in the era of HAART because more unprotected anal sex is
likely to take place between steady couples other that casual ones,
according to
\cite{Baggaley2005,Xiridou2003,Xiridou2004,Semple2003,Davidovich2004}.
Thus the fraction of steady relationships in a network is regarded
as an important parameter. Based on questionnaire results given by
\cite{Xiridou2003}, approximately 50\% of the ACS participants
reported having a steady partner. So we estimate that a partnership
is initialized as a steady one with a probability given by

\begin{equation} \label{Psteady}
P_{steady}= N*(0.5/2)/\left<e\right>
\end{equation}

where
 \begin{eqnarray}
%\begin{split}
\left<k\right>&=&\sum^{k_{max}}_{k=0}kp_{k}, \nonumber\\
\left<e\right>&=& N*\left<k\right>/2 \nonumber
% \end{split}
 \end{eqnarray}

$N$ is the size of a network, $\left<e\right>$ is the expected total
number of edges in a network and $\left<k\right>$ is the expected
degree of vertices. According to Equation \ref{Pk} with given values
of $\gamma$ and $k_{max}$, we get $P_{steady}\approx0.054$. Allowing
for steady partnerships being kept until expired, the averaged
fraction of steady relationships from our simulation results
fluctuates between 0.047 and 0.055.

The duration of steady partnerships among Amsterdam MSM is reported
to have an expected value of 1.5 years \cite{Xiridou2003}. Thus we
assume the duration of steady partnerships following a discrete
uniform distribution $\mathcal{DU}(1,2)$.

Unlike the PI or AIDS stages with fixed durations, the duration of
the AP stage can vary with the therapy effecitiveness. So we draw
the duration of AP from a Poisson distribution with a mean value of
13 years for a failed treatment, and a mean value of 22 years for a
successful treatment \cite{Xiridou2003}.

The transmission probability for sexual actions other than
UAI\footnote{\textbf{UAI}, unprotected anal intercourse;
\textbf{URAI}, unprotected receptive anal intercourse;
\textbf{UIAI}, unprotected insertive anal intercourse} is assumed to
be negligible\cite{Caceres1994}, also, the transmission probability
for an UAI between serodiscordant\footnote{\textbf{Serodiscordant}
(sero-discordant) is a term used to describe a couple in which one
partner is HIV positive and the other is HIV negative (Smith,
Raymond. Couples. Retrieved on 08-26, 2006.).
\textbf{Seroconcordant} is the term used to describe a couple in
which both partners are of the same HIV status (ie both are HIV
positive or both are HIV negative).} men is depending on whether the
role of the HIV-negative man is receptive (URAI) or insertive
(UIAI). We assume each partner of a couple takes randomly a
receptive or insertive role in an action. The frequency of either
URAI or UIAI between steady partners is 15 per
year\cite{Xiridou2003}, for simplicity we assume that the frequency
of sexual actions of MSM per year follows a Poisson distribution of
$\mathcal{P}(30)$. At the PI stage, the frequency within the first 3
months and the last 9 months is assumed to follow $\mathcal{P}(8)$
and $\mathcal{P}(22)$, respectively.

\subsubsection{On an individual scale}
Epidemic parameters and the involved reduction factors are
discussed.

\

Of the infected individuals at the asymptomatic stage, 42\% know
they are HIV positive\cite{Xiridou2003}. Since the introduction of
HAART in 1996, the fraction of individuals that has received any
form of HAART therapy after diagnosis, is 81\% in 2006
\cite{ACS2006}. 70\% of the individuals who received therapy will be
successfully treated\cite{Xiridou2003}. In other words, individuals
get diagnosed with a probability of 0.42, and obtain a treatment
with a probability of 0.81, and then get successfully treated with a
probability of 0.7.

The value of $TP_{i}$ (see Section \ref{propagation}) is related to
the infection stage of an individual $V_{i}$, given by
$TP_{\scriptsize{\textrm{PI}},1}$,
$TP_{\scriptsize{\textrm{PI}},2}$, $TP_{\scriptsize{\textrm{AP}}}$
or $TP_{\scriptsize{\textrm{AIDS}}}$. According to
\cite{Leynaert1998,Vittinghoff1999,DeGruttola1989}, the probability
of the transmission per URAI/UIAI act at stage PI and AP is
0.22/0.044 and 0.011/0.0022 respectively. Therefore,
$TP_{\scriptsize{\textrm{PI}},1} = 0.22/0.044$ and
$TP_{\scriptsize{\textrm{PI}},2} = TP_{\scriptsize{\textrm{AP}}}=
0.011/0.0022$ with respect to the receptive or insertive role.

ART is assumed to reduce the transmissibility by 50-90\% to a
moderate extent in \cite{Boily2004}, so that in our simulations the
treatment factor $F_{T_{i}}$ is set to follow a continuous uniform
distribution $\mathcal{CU}(0.1,0.5)$. If an individual has a steady
partner, the partnership factor $F_{P_{i}}$ is 0.84\citep[see][Sec.
2]{Xiridou2003}, otherwise 1 which means no reduction.

The general use of ART leading to changes in risk behavior has
sparked considerable concerns, for instance, Baggaley discussed in
\cite{Baggaley2005} that increases in risk behavior could result
from increased optimism about HIV therapy due to the availability of
HAART. In our simulations, the definition and estimated values of
risk behavior factors $F_{R_{ij}}$ is based on
\cite{Dukers2001,Bezemer2008}, and the relative values are given in
Table \ref{riskfactor}.

\begin{table}
\centering \caption{The yearly relative values of risk behavior
factor for HIV negative and positive MSM. * - The value in 1987 is
an average of its adjacent years (1986 and 1988), so is that in
1996. ** - The baseline with value 1.00.} \label{riskfactor}
\begin{tabular}{lccccccc}\hline

  & {1985-1986} & {1987} & {1988-1991} & {1992-1995} & {1996} &
{1997-1999} & {2000-}  \\\hline

negative & 3.50 & $2.50^*$ & 1.50 & 0.80 & 0.90 &
\textbf{$1.00^{**}$}
& 1.30\\

positive & 2.80 & 1.61 & 0.42 & 0.88 & 0.78 & 0.70 & 1.30\\\hline

%\end{tabularx}
\end{tabular}
\end{table}

\subsection{Results and discussion}
\begin{figure}
%\centering

 \caption{Simulation results (averaged on 24 executions) versus statistical historical data of HIV incidence (ACS)(a) and AIDS diagnoses cases (in Netherlands, not only in Amsterdam)(b) over calendar years}
\label{resultFig}

\subfigure[Incidence]{
\includegraphics[width=200 bp]{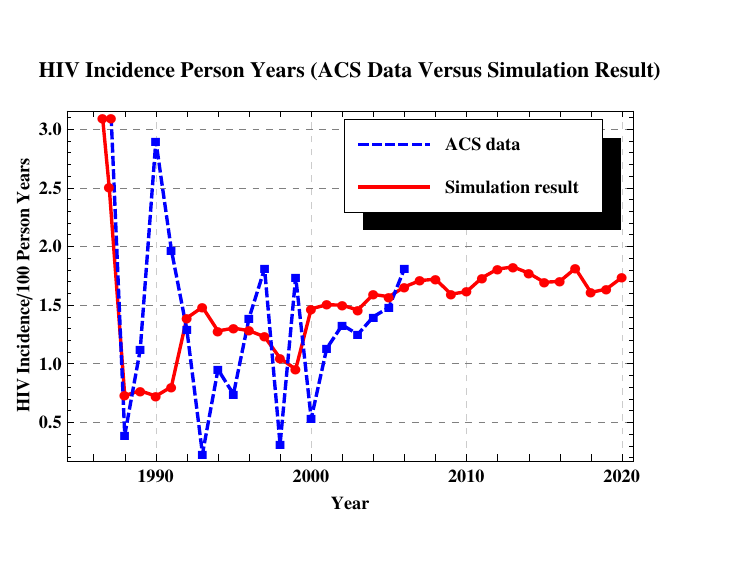}}
\subfigure[AIDS Diagnoses]{
\includegraphics[width=200 bp]{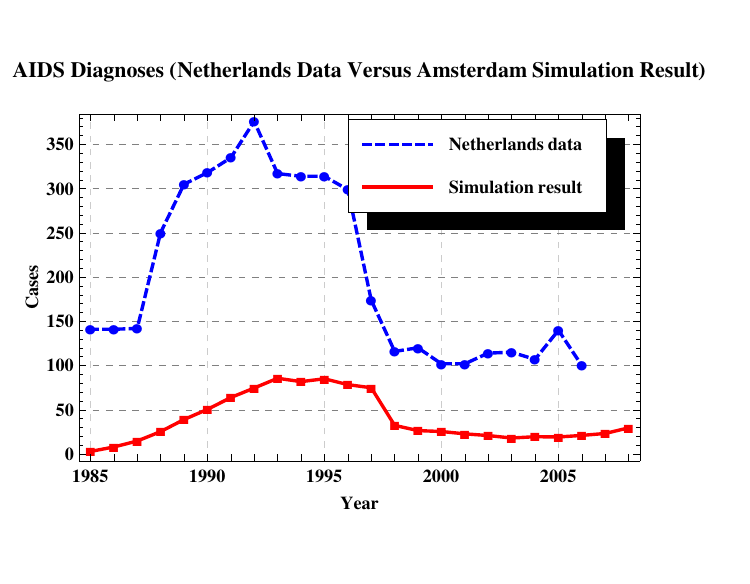}}
\end{figure}

The simulation results illustrated in Figure \ref{resultFig} shows
the HIV incidences (a) and AIDS diagnosis cases (b) over calendar
years.

We used mainly the incidence for comparison and validation because
of its relative dimension. Our simulated yearly incidences are
consistent with the historical ACS data \citep[see][Table
5]{ACSOverview2006}, based on the conduct of a null hypothesis test
where the simulation result of incidence does not differ
significantly from the ACS data. Although the historical data (the
dashed line in Figure 4a), got based on incomplete statistics (due
to the involvement of subgroups of MSM related to substudies of ACS
in years), was fluctuating tremendously, the hypothesis is accepted
by using Chi-Square tests with a significance level of 0.05 (data
not shown).

As a supplement, we compared the simulated AIDS diagnosis cases with
the historical statistics in the Netherlands in the aspect of trend.
Our result shows, even that it is based on a relatively small sample
space with 2299 Amsterdam MSM, a similar trend to the statistical
data in the Netherlands (including Amsterdam as one city) over
calendar years \citep[see][Table B.18]{Veen2006}. Both the two
curves in Figure 4b approximately show a continuous increase in AIDS
diagnosis cases until 1996 and an acute decline in 1996-1998, and
then keeps stable from 1999 on. In this study, our simulation
suggests that the acute decline is most likely caused by the
introduction of HAART in 1996. In the future, we could scale up the
simulation of AIDS diagnosis cases from Amsterdam to the
Netherlands, if and when the total numbers of MSM with different
behavior features in different cities of Netherlands are available.

\section{Conclusions}
\label{conclusion}

We proposed a flexible hybrid Complex Agent Network approach and
validated experimentally its applicability by taking the dynamics of
HIV propagation among Amsterdam MSM, a representative small
population, as a case study. Our experiments showed good
correspondence between the model results and the historical data of
ACS, in addition, the results are robust with respect to small
changes in the input parameters (data not shown). Therefore, the
model can be adopted to predict the future trend of HIV prevalence
among MSM in Amsterdam, whilst allowing for easy tuning the values
of highly concerned parameters, such as the drug effectivity,
diagnosis proportion and risk behavior factor.

\section{Acknowledgments}
The authors would like to acknowledge the financial support of the
China Scholarship Council (www.csc.edu.cn) and the European ViroLab
\cite{ViroLabSloot2008} (www.virolab.org) grant INFSO-IST-027446.

\end{document}